\documentclass[aps,twocolumn,superscriptaddress,pra]{revtex4}
\usepackage{graphicx,amsmath,color}
\usepackage{bm}
\usepackage{amsfonts}

\begin{document}

\title{Moving obstacle potential in a spin-orbit-coupled Bose-Einstein condensate}

\author{Masaya Kato}
\affiliation{Department of Engineering Science, University of Electro-Communications, Tokyo 182-8585, Japan}

\author{Xiao-Fei Zhang}
\affiliation {Key Laboratory of Time and Frequency Primary Standards, National Time Service Center, Chinese Academy of Sciences, Xi'an 710600, China}
\affiliation {University of Chinese Academy of Sciences, Beijing 100049, China}

\author{Hiroki Saito}
\affiliation{Department of Engineering Science, University of Electro-Communications, Tokyo 182-8585, Japan}

\date{\today}
\begin{abstract}

We investigate the dynamics around an obstacle potential moving in the
plane-wave state of a pseudospin-$1/2$ Bose-Einstein condensate with
Rashba spin-orbit coupling.
We numerically investigate the dynamics of the system and find that it
depends not only on the velocity of the obstacle but also significantly on
the direction of obstacle motion, which are verified by a Bogoliubov
analysis.
The excitation diagram with respect to the velocity and direction is
obtained.
The dependence of the critical velocity on the strength of the spin-orbit
coupling and the size of the obstacle is also investigated.
\end{abstract}


\maketitle

\section{Introduction}
\label{s:introduction}

The successful experimental realization of various synthetic gauge fields and
spin-orbit coupling (SOC) in Bose-Einstein condensates (BECs) has recently
drawn considerable theoretical attention~\cite{Y.-J. Lin1,Y.-J. Lin2,Y.-J. Lin3,Y.-J. Lin4,
B. M. Anderson,Z. Fu,S. C. Ji0,L. Huang,Z. M. Meng,Z. Wu}, with a number
of studies 
addressing the ground state structures and static properties of topological
excitations (for recent reviews, see, for example,
Refs.~\cite{J. Dalibard,N. Goldman,H. Zhai0}).
Such synthetic gauge potentials in cold atoms are powerful tools for
quantum many-body simulators of real materials.
Moreover, the spin-orbit-coupled (SO-coupled) BEC exhibits numerous novel
phases that cannot be found in conventional condensed matter systems.

In the present paper, we focus on the problem of the moving obstacle
potential in an SO-coupled BEC.
The drag force on a moving impurity in a SO-coupled BEC has been
calculated using the Bogoliubov spectrum~\cite{P.-S. He,R. Liao}.
In contrast, we directly solve the Gross-Pitaevskii (GP) equation with SOC
and investigate the dynamical effects of the moving obstacle potential on
the SO-coupled condensate.
Most theoretical studies on BEC with Rashba SOC have focused on the static
properties of the
condensate~\cite{C. Wang,T. L. Ho,X.-F. Zhou,S. Sinha,H. Hu,T. Kawakami,B. Ramachandhran},
and there have only been a few studies on dynamics~\cite{J. Radic,A. L. Fetter, K. Kasamatsu,A. Gallem}.

The Dynamics of an SO-coupled BEC around a moving obstacle potential
differs significantly from that of the usual scalar BEC in two ways.
First, the ground state of the SO-coupled BEC breaks rotational symmetry,
and the excitation spectrum above the ground state has an anisotropic
characteristic form with a roton minimum~\cite{Q. Zhu,M. A. Khamehchi,
S. C. Ji,K. Sun}.
Because of this feature, the Landau critical velocity and excitation
properties depend on the direction of obstacle motion.
Second, due to the close relationship between the spin and motional
degrees of freedom, the dynamic properties of quantized vortices and
solitons are dramatically altered by the SOC~\cite{M. Kato}.
Consequently, the generation of vortices and waves around the moving
obstacle potential is significantly affected by the SOC.

The remainder of the present paper is organized as follows. In
Sec.~\ref{s:formulation}, we formulate the theoretical model
for a moving obstacle potential in a uniform SO-coupled BEC. 
In Sec.~\ref{s:dynamics},
the excitation dynamics induced by the obstacle potential and the
Bogoliubov analysis are presented.
The parameter dependence and velocity field are investigated in
Secs.~\ref{s:parameter} and \ref{s:VF}, respectively.
Finally, in Sec.~\ref{s:conclusions}, the main results of the present
paper are summarized.

\section{Formulation of the problem}
\label{s:formulation}
We consider a BEC of quasispin-$1/2$ atoms with Rashba SOC, where an
obstacle potential is moving in a uniform system.
The mean-field approximation is used, and the dynamics of the system is
described by the GP equation as follows:

\begin{eqnarray}
\label{eq:GP}
i \hbar \frac{\partial}{\partial t}\bm \Psi =
- \frac{\hbar^2}{2m}\bm{\nabla}^2\bm{\Psi}
&+&i\frac{\hbar k_0}{m} \bm{\nabla} \cdot \bm{\sigma}_\perp \bm\Psi
\nonumber\\
&+&U(\bm r,t)\bm\Psi
 + \hat{G}\left( \bm\Psi,\bm \Psi^\dagger\right)\bm\Psi,
\end{eqnarray}
where $\bm\Psi(\bm r)=(\psi_{1}(\bm r),\psi_{2}(\bm r))^T$ is the spinor order parameter,
$m$ is the mass of atoms, $k_0$ is the SOC coefficient,
$\bm{\sigma}_\perp = (\sigma_x, \sigma_y, 0)$ are $2\times 2$ Pauli matrices,
and $U(\bm r,t)$ is a moving obstacle potential.
The interaction matrix in Eq.~(\ref{eq:GP}) is given by
\begin{equation}
\hat{G}(\bm\Psi,\bm\Psi^\dagger) =
\begin{pmatrix}
\textmd{g}_0|\psi_1|^2 &\textmd{g}_{12}|\psi_2|^2\\
\textmd{g}_{12}|\psi_1|^2 &\textmd{g}_0|\psi_2|^2
\end{pmatrix},
\end{equation}
where $\textmd{g}_0$ and $\textmd{g}_{12}$ are the intra- and
inter-component interaction coefficients, respectively. (Here, we further
assume that the two intracomponent interaction parameters are the same.)
We consider an infinite system in which the atomic density
$\bm\Psi^\dagger\bm\Psi$ far from the potential is constant, $n_0$. In the
following, we normalize the length and time by the healing length
$\xi=\hbar/(\textmd{g}_0 n_0 m)$ and the characteristic time scale $\tau =
\hbar/(\textmd{g}_0 n_0)$.
In this unit, the wave function, velocity, and energy are normalized by
$n_0$, $\sqrt{\textmd{g}_0 n_0/m}$, and $\textmd{g}_0 n_0$.
We transform Eq.~(\ref{eq:GP}) into a frame of reference that moves with
the potential at velocity $\bm v=(v \cos \theta_v, v \sin \theta_v)$.
The normalized GP equation becomes
\begin{subequations}
\label{eq:GPE}
\begin{eqnarray}
i\frac{\partial\psi_1}{\partial t} =
-\frac{1}{2}\bm{\nabla}^2\psi_1
+i\kappa\partial_{-}\psi_2
+i\bm v \cdot \bm\nabla\psi_1 \nonumber\\
+V(\bm r)\psi_1
+\left(|\psi_1|^2+\gamma|\psi_2|^2 \right)\psi_1,\\
i\frac{\partial\psi_2}{\partial t} =
-\frac{1}{2}\bm{\nabla}^2\psi_2
+i\kappa\partial_{-}\psi_1
+i\bm v \cdot \bm\nabla\psi_2 \nonumber\\
+V(\bm r)\psi_2
+\left(\gamma|\psi_1|^2+|\psi_2|^2 \right)\psi_2,
\end{eqnarray}
\end{subequations}
where $\partial_{\pm}=\partial/\partial x \pm i \partial/\partial y$, $\kappa = k_0/\sqrt{g_0 n_0 m}$, and
$\gamma = \textmd{g}_{12}/\textmd{g}_0$.
We use a circular potential with radius $R$ as
\begin{equation}
V(\bm r)=
\begin{cases}
V_0 & (|\bm r| \leq R)\\
0 & (|\bm r| > R),
\end{cases}
\end{equation}
where the potential height $V_0$ is taken to be much larger than the
chemical potential.

The ground state without the potential is the plane-wave state for
$\gamma<1$ and the stripe state for $\gamma > 1$.
In the following discussion, we focus on the miscible case, $\gamma<1$,
and the ground state far from the potential is given by the plane-wave
state, as follows:
\begin{equation}
\label{eq:PL}
\bm \Psi(\bm r) =
\frac{1}{\sqrt{2}}
\left(
\begin{array}{c}
e^{i \kappa x}\\
e^{i \kappa x}
\end{array}
\right),
\end{equation}
where the wave vector is chosen to be in the $x$ direction.

We numerically solve Eq.~(\ref{eq:GPE}) by the pseudospectral method using
the fourth-order Runge-Kutta scheme.
The initial state is the ground state with $\bm{v} = 0$, which is the
plane-wave state in Eq.~(\ref{eq:PL}) far from the potential.
The initial state is prepared by the imaginary-time propagation method, in
which $i$ on the left-hand side of Eq.~(\ref{eq:GPE}) is replaced with
$-1$.
The numerical space is taken to be $400 \times 400$, which is sufficiently large, and the
effect of the periodic boundary condition can be neglected.

\section{Excitation induced by an obstacle} \label{s:dynamics}

\subsection{Dynamics of the system}

\begin{figure}[htb]
\includegraphics[width=9cm]{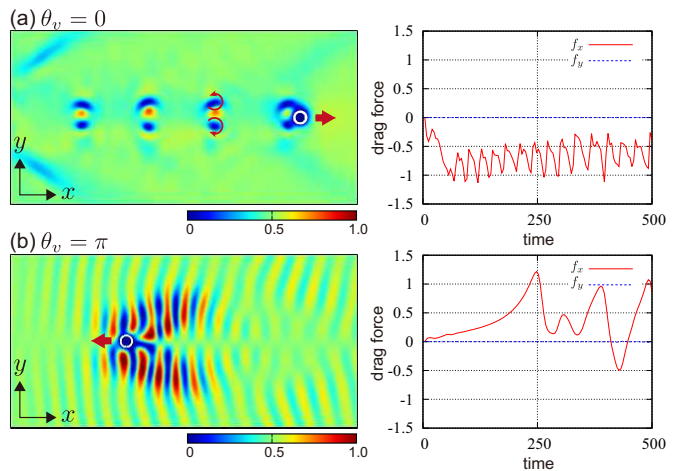}
\caption{
Typical snapshots of the density distribution $|\psi_1|^2$ and time evolution of
the drag force $\bm{f}$ exerted on the potential.
The velocities of the potential $(v,\theta_v)$ are
(a) $(0.5,0)$ and (b) $(0.05,\pi)$.
The field of view of the left-hand panels is $60 \times 30$.
The white circles represent the obstacle potential, and the thick arrows
indicate the moving direction.
The curved arrows indicate the directions of the circulations of the
vortices.
The parameters are $R=0.5$, $\kappa=1$, and $\gamma=0.8$.
See the Supplemental Material for movies of the dynamics of
component 1~\cite{SM}.
}
\label{f:transverse}
\end{figure}
First, we focus on the two special cases of $\theta_v = 0$ and
$\theta_v = \pi$, where $\theta_v$ is the angle between the obstacle
velocity $\bm{v}$ and the $x$ axis.
In the case of $\theta_v=0$, when the velocity $v$ exceeds a critical
value, vortex-antivortex pairs are created, as shown in the left-hand
panel of Fig.~\ref{f:transverse}(a). 
In this case, the periodic generation of vortex-antivortex pairs is, in a
sense, reminiscent of the scalar case.
However, the created vortex pairs are different from those in the scalar
BEC, in that the vortex cores in both components deviate from each other,
producing pairs of half-quantum vortices~\cite{M. Kato}.
The right-hand panel of Fig.~\ref{f:transverse}(a) shows the drag force
experienced by the obstacle, defined by $\bm f_{\rm d}
= i \partial_t \int d \bm r (\Psi^{\dagger} \bm{\nabla} \Psi)$.
The drag force in the $x$ direction exhibits periodic oscillation due to
the periodic generation of vortices, whereas the drag force in the $y$
direction remains zero.
In the case of $\theta_v = \pi$, when the velocity $v$ exceeds a critical
velocity, spin waves are excited, as shown in the left-hand panel of
Fig.~\ref{f:transverse}(b), which is very different from the case of
$\theta_v=0$.
In this spin-wave state, high-density regions of components 1 and 2 are
alternately aligned, as in the stripe state of an SOC BEC.
The critical velocity of the spin-wave generation is much smaller than
that of the vortex generation for $\theta_v=0$.
Thus, the excitation dynamics is strongly anisotropic with respect to the
moving obstacle potential.

\begin{figure}[t!]
\includegraphics[width=8cm]{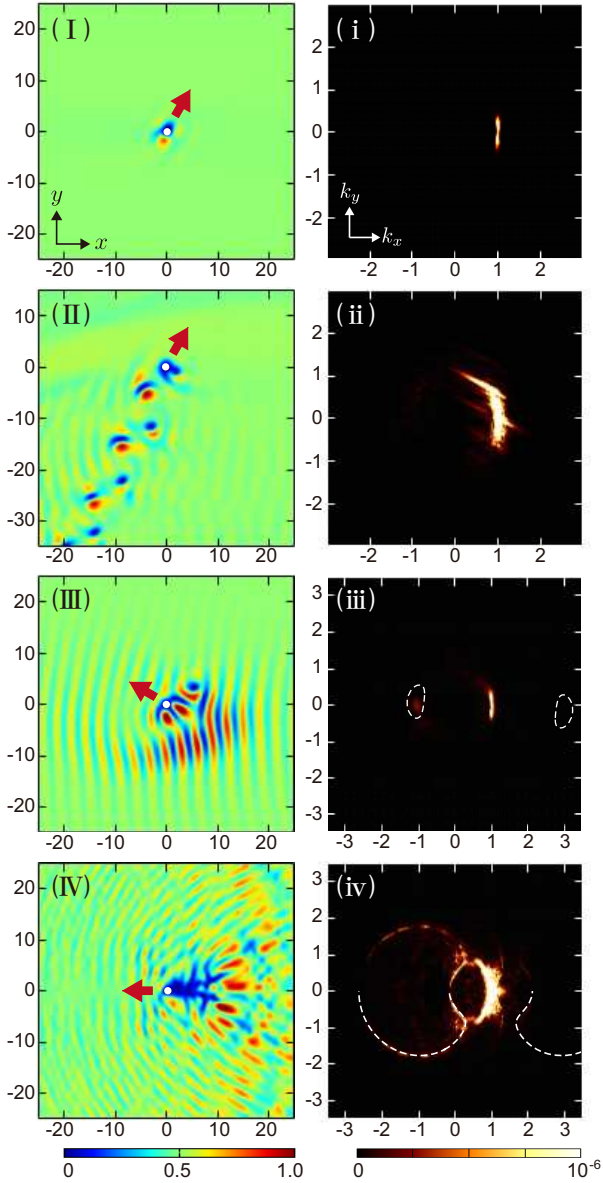}
\caption{
Snapshots of (I)-(IV) the density distributions $|\psi_1|^2$ and (i)-(iv)
the momentum-space distribution $|\tilde{\psi}_1|^2$ for $\kappa=1$,
$\gamma=0.8$, and $R = 0.5$, where $\tilde{\psi}_1$ is the Fourier
transform of $\psi_1$.
(I, i) $(v,\theta_v)=(0.1,\pi/3)$, (II, ii) $(v,\theta_v)=(0.2,\pi/3)$,
(III, iii) $(v,\theta_v)=(0.07,5\pi/6)$,
and (IV, iv) $(v,\theta_v)=(0.4,\pi)$.
The white solid circles and the red arrows in the left-hand panels are the obstacle potentials and moving directions, respectively.
The dashed lines in (iii) and (iv) indicate the analytical solutions of
$\omega(k_x - \kappa, k_y)=0$ in Eq.~(\ref{eq:det_B}).
(In (iv), the dashed lines are shown only for $k_y < 0$.)
See the Supplemental Material for movies of the dynamics of
$|\psi_1|^2$ and $|\tilde{\psi}_1|^2$~\cite{SM}.
}
\label{f:dynamics}
\end{figure}
For a deeper understanding of the anisotropic properties, we explore the
$v$ and $\theta_v$ dependence of the dynamics.
Figure~\ref{f:dynamics} shows four typical dynamics with different
velocities and azimuthal angles of the obstacle motion.
These four kinds of dynamics are (I) weak excitation, (II) vortex pairs,
(III) spin wave, and (IV) strong excitation.
In the absence of excitation, the momentum-space distribution is
$\delta(k_x-\kappa, k_y)$ due to the plane-wave background.
For the weak excitation, no pronounced excitations are observed in the
density distributions, but the long wavelength excitations are induced at
$k_x = \kappa$ and $|k_y| \lesssim 0.4$, as shown in
Fig.~\ref{f:dynamics}(i).
For the spin wave, we find excitations at $\bm k \simeq -\kappa \bm e_x$,
as shown in Fig.~\ref{f:dynamics}(iii).
For the strong excitation, the shock-wave pattern appears in the density
distribution, as shown in Fig.~\ref{f:dynamics}(IV), which has a ring
shape in the momentum space, as shown in Fig.~\ref{f:dynamics}(iv).
These momentum-space behaviors are explained in Sec.~\ref{s:bogoliubov}.

\begin{figure}[htb]
\includegraphics[width=9cm]{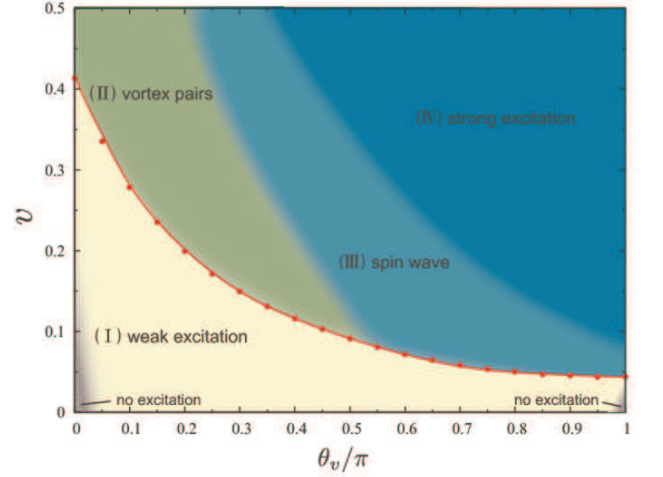}
\caption{
The excitation diagram with respect to $v$ and $\theta_v$ for $\kappa=1$,
$\gamma=0.8$, and $R=0.5$.
The dynamics are classified into (I)-(IV), which correspond to those in 
Fig.~\ref{f:diagram}.
The red points indicate the critical velocity, at which the drag force
increases suddenly, and the red line is a visual guide.
}
\label{f:diagram}
\end{figure}
Figure~\ref{f:diagram} shows a diagram of the four kinds of excitation
with respect to $v$ and $\theta_v$, where regions (I)-(IV) correspond
to the dynamics in Fig.~\ref{f:dynamics}.
As $v$ and $\theta_v$ are increased, the excitation behavior changes from
(II) to (IV).
The boundaries between these three regions are vague.
In contrast, there is a sharp boundary between region (I) and regions
(II)-(IV), as indicated by the red points in Fig.~\ref{f:diagram}.
When we start from region (I) and slowly increase the velocity $v$,
the small drag force in region (I) steeply increases at this boundary.
For small $v$ and $\theta \simeq 0, \pi$, there are narrow regions in
which the drag force almost vanishes, which are indicated by the dark
regions in Fig.~\ref{f:diagram}.

\subsection{Bogoliubov analysis}
\label{s:bogoliubov}

\begin{figure*}[tb]
\includegraphics[width=18cm]{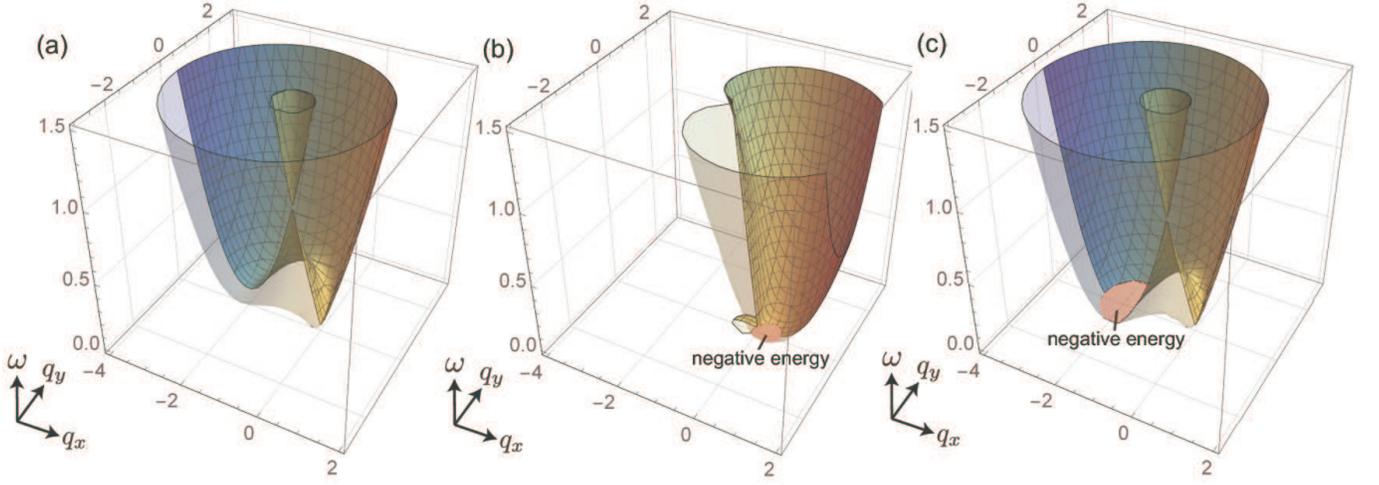}
\caption{
Excitation spectrum of a uniform system for $\kappa=1$ and $\gamma=0.8$ in
the frame moving at velocities (a) $(v,\theta_v)=(0,0)$, (b)
$(v,\theta_v)=(1,0)$, and (c) $(v,\theta_v)=(0.1,\pi)$.
The colored regions on the $q_x$-$q_y$ plane indicate negative energy,
$\omega(\bm q)<0$.
}
\label{f:bogoliubov}
\end{figure*}
We perform a Bogoliubov analysis in order to clarify the numerically
obtained behavior.
The wave function can be written as
\begin{equation}
\label{eq:VWF}
\bm\Psi(\bm r,t)=
e^{-i\mu t + i\kappa x}
\left[
\frac{1}{\sqrt{2}}
\left(
\begin{array}{c}
1\\
1
\end{array}
\right)
+\bm\delta \bm\Psi(\bm r,t)
\right],
\end{equation}
where $\mu = -\kappa^2/2- v_x\kappa +(1+\gamma)/2$ is the chemical
potential.
The small excitation $\delta \bm\Psi(\bm r,t)$ is decomposed into
\begin{equation}
\label{eq:fl}
\delta \bm\Psi(\bm r,t)=
\bm\alpha e^{i(\bm q \cdot \bm r -\omega t)}+\bm\beta^{\ast} e^{-i(\bm q \cdot \bm r-\omega t)},
\end{equation}
where $\bm\alpha=(\alpha_1,\alpha_2)^{\rm T}$ and
$\bm\beta=(\beta_1,\beta_2)^{\rm T}$ are the amplitudes,
$\bm q=(q\cos\theta_q,q\sin\theta_q)$ is the wave number, and 
$\omega$ is the frequency of the excitation.
Substituting Eq.~(\ref{eq:VWF}) into Eq.~(\ref{eq:GPE}) with $V = 0$
and taking the first order of $\delta\bm\Psi$, we obtain
\begin{equation}
\label{eq:BE_fl}
\left(\mu' + i \frac{\partial}{\partial t}\right) \delta \bm{\Psi} =
(\mathcal{H}_0 + \mathcal{H}_1)\delta\bm{\Psi}+\mathcal{H}_2 \delta\bm{\Psi}^\ast + i \bm v \cdot \nabla \delta \bm{\Psi},
\end{equation}
where $\mu' = -\kappa^2/2 +(1+\gamma)/2$,
\begin{equation}
\mathcal{H}_0 =
\begin{pmatrix}
-\frac{1}{2}(\bm\nabla + i\kappa\bm e_x)^2 & i\kappa\partial_- - \kappa^2\\
i\kappa\partial_+ - \kappa^2& -\frac{1}{2}(\bm\nabla + i\kappa\bm e_x)^2
\end{pmatrix},
\end{equation}
and
\begin{equation} \label{H12}
\mathcal{H}_1 = \frac{1}{2}
\begin{pmatrix}
2+\gamma & \gamma\\
\gamma & 2+\gamma
\end{pmatrix},
\mathcal{H}_2 = \frac{1}{2}
\begin{pmatrix}
1 & \gamma \\
\gamma & 1
\end{pmatrix}.
\end{equation}
The Bogoliubov equation is obtained from Eqs.~(\ref{eq:fl})-(\ref{H12}) as
$\{\mathcal{H}(\bm q) - [\omega(\bm q) + \bm v \cdot \bm q]\mathcal{I}\}\bm \chi = \bm 0$, where
\begin{widetext}
\begin{equation}
\label{eq:BH}
\mathcal{H}(\bm q) =
\begin{pmatrix}
\frac{1}{2}q^2+\kappa q_x +\kappa^2 + \frac{1}{2} &
-\kappa q_{-} - \kappa^2 + \frac{1}{2}\gamma  &
\frac{1}{2} & \frac{1}{2}\gamma  \\
-\kappa q_{+} - \kappa^2 + \frac{1}{2}\gamma  &
\frac{1}{2}q^2+\kappa q_x +\kappa^2 + \frac{1}{2} &
\frac{1}{2}\gamma  & \frac{1}{2}  \\
-\frac{1}{2} & -\frac{1}{2}\gamma  &
-\frac{1}{2}q^2+\kappa q_x -\kappa^2 - \frac{1}{2} &
-\kappa q_{+} + \kappa^2 - \frac{1}{2}\gamma  \\
-\frac{1}{2}\gamma  & -\frac{1}{2}  &
-\kappa q_{-} + \kappa^2 - \frac{1}{2}\gamma  &
-\frac{1}{2}q^2+\kappa q_x -\kappa^2 - \frac{1}{2}
\end{pmatrix}
\end{equation}
\end{widetext}
and $\bm \chi=(\alpha_1,\alpha_2, \beta_1,\beta_2)^{\rm T}$.
The Bogoliubov excitation spectrum $\omega(\bm q, \bm v)$ is the solution
of
\begin{equation}
\label{eq:det_B}
|\mathcal{H}(\bm q) - (\omega + \bm v \cdot \bm q)\mathcal{I}|=0.
\end{equation}

Figure~\ref{f:bogoliubov}(a) shows the excitation spectrum
$\omega(\bm q, \bm v)$ at $\bm{v} = \bm{0}$.
The spectrum breaks the rotational symmetry and has a roton-like minimum at
$q_y < 0$.
For nonzero $v$, a region in which $\omega$ becomes negative appears.
The Landau critical velocity is defined as the velocity above which the
negative-$\omega$ region appears on the $q_x$-$q_y$ plane.
Thus, in the present case, the Landau critical velocity depends on the
angle $\theta_v$ of the moving potential.
When $\theta_v = 0$, the negative-energy region is located near the origin
of the momentum space, as shown in Fig.~\ref{f:bogoliubov}(b), which is
similar to the case of a scalar BEC.
For $\theta_v = \pi$, on the other hand, the negative-energy region
appears at finite momentum due to the roton-like minimum, as shown in
Fig.~\ref{f:bogoliubov}(c).
This corresponds to the spin-wave excitation in
Figs.~\ref{f:dynamics}(III) and \ref{f:dynamics}(iii).

Setting $\omega=0$ in Eq.~(\ref{eq:det_B}), we have
\begin{equation}
\label{eq:det}
|\mathcal{H}(\bm q) - \bm v \cdot \bm q \mathcal{I}|=0.
\end{equation}
For given $\theta_v$, the Landau critical velocity $v_L$ is the minimum
value of $v \geq 0$ for which Eq.~(\ref{eq:det}) has at least a real
solution $\bm{q}$.
First, we solve Eq.~(\ref{eq:det}) under the assumption that the solution
is $q \simeq 0$.
Taking the limit $q \rightarrow 0$, Eq.~(\ref{eq:det}) is rewritten as
\begin{equation}
 \label{eq:vs_q0}
 v=\sqrt{\frac{1+\gamma}{2}}\left|\frac{\cos\theta_q}{\cos(\theta_q - \theta_v)}\right|,
\end{equation}
where $\theta_q$ is the azimuthal angle in the $q_x$-$q_y$ plane.
When $\theta_v \neq 0$ and $\theta_v \neq \pi$, the minimum value of $v
\geq 0$ for which Eq.~(\ref{eq:vs_q0}) has a solution is $v = 0$ (the
solution is $\theta_q = \pm \pi / 2$).
Thus, the Landau critical velocity for $\theta_v \neq 0$ and $\theta_v
\neq \pi$ is $v_L = 0$.
The instability around $\theta_q = \pm \pi / 2$ appears in
Fig.~\ref{f:dynamics}(i).
The Landau critical velocities for $\theta_v = 0$ and $\theta_v = \pi$ are
obtained analytically, and we have
\begin{equation}
\label{eq:SV}
 v_L(\theta_v)=
  \begin{cases}
    \sqrt{(1+\gamma)/2} & (\theta_v = 0)\\
    -\kappa + \sqrt{\kappa^2 + (1-\gamma)/2} & (\theta_v = \pi) \\
    0 & ({\rm other})
  \end{cases}.
\end{equation}
Note that $v_L$ for $\theta_v = 0$ is independent of $\kappa$, which
has a phonon-like relation $\omega \propto q_x$ for $q_x \ll 1$, as in the
Bogoliubov mode of a scalar BEC.
When $\theta_v = \pi$, the wave number
\begin{equation} \label{qx}
q_x = -\sqrt[4]{8\kappa^2(1-\gamma+2\kappa^2)}
\end{equation}
first becomes negative
above the Landau critical velocity, which corresponds to the wave number
for the spin wave shown in Figs.~\ref{f:dynamics}(III) and
\ref{f:dynamics}(iii).
In the limit of $\kappa \rightarrow 0$, the two velocities in
Eq.~(\ref{eq:SV}) are $\sqrt{(1 \pm \gamma) / 2}$, which agree with those
in the two-component BEC without SO coupling.

Although the Landau critical velocity is 0 for $\theta_v \neq 0$ and
$\theta_v \neq \pi$ in Eq.~(\ref{eq:SV}), the effect of the excitation
remains slight when $v$ is small, as shown in Figs.~\ref{f:dynamics}(I)
and \ref{f:dynamics}(i).
However, there exists an effective critical velocity, above which the drag
force suddenly increases, as indicated by the red line in
Fig.~\ref{f:diagram}.

The dashed lines in Figs.~\ref{f:dynamics}(iii) and \ref{f:dynamics}(iv)
indicate the analytical solutions of $\omega(k_x - \kappa, k_y, \bm v)=0$
for $(v, \theta_v)=(0.07, 5\pi/6)$ and $(v, \theta_v)=(0.4, \pi)$,
respectively.
In Fig.~\ref{f:dynamics}(iii), condensates are excited in the region
of $\omega(k_x - \kappa, k_y, \bm v)<0$.
For the much larger velocity, a ring excitation appears in the momentum
space, which we classified into (IV) strong excitation.
Due to the energy conservation law, the entire region of $\omega(k_x -
\kappa, k_y, \bm v)<0$ inside the dashed line in
Fig.~\ref{f:dynamics}(iv) cannot be excited, and only the ring-like
region at $\omega(k_x - \kappa, k_y, \bm v)\simeq 0$ is excited.

For $\theta_v=0$ and $\theta_v=\pi$, Eq.~(\ref{eq:BH}) can be
diagonalized, and the eigenvectors can be obtained.
For $\theta_v=0$, the eigenvector is
\begin{subequations}
 \label{eq:EV_0}
  \begin{eqnarray}
   \bm \alpha(q) \propto \sqrt{1+f(q)}
    \left(
     \begin{array}{c}
      1\\
      1
     \end{array}
    \right),\\
   \bm \beta(q) \propto -\sqrt{1-f(q)}
    \left(
     \begin{array}{c}
      1\\
      1
     \end{array}
    \right),
  \end{eqnarray}
\end{subequations}
where
\begin{equation}
 f(q)=\frac{q\sqrt{q^2 + 2\left(1 + \gamma \right)}}{q^2 + 1 + \gamma},
\end{equation}
and for $\theta_v=\pi$,
\begin{subequations}
 \label{eq:EV_pi}
  \begin{eqnarray}
   \bm \alpha(q) \propto \sqrt{1+g(q)}
    \left(
     \begin{array}{c}
      1\\
      -1
     \end{array}
    \right),\\
   \bm \beta(q) \propto \sqrt{1-g(q)}
    \left(
     \begin{array}{c}
      -1\\
      1
     \end{array}
    \right),
  \end{eqnarray}
\end{subequations}
where
\begin{equation}
 g(q)=\frac{\sqrt{\left(q^2 + 4\kappa^2\right) \left(q^2 + 4\kappa^2 + 2  - 2\gamma\right)}}{q^2 + 4\kappa^2 + 1 - \gamma}.
\end{equation}
In the case of $\gamma \simeq 1$, we find $|\bm \beta|/|\bm \alpha| \simeq
0$ in $\theta_v = \pi$.
This is why the Bogoliubov counterparts (right-hand dashed
lines in Figs.~\ref{f:dynamics}(iii) and \ref{f:dynamics}(vi)) are not
significantly excited.

\section{Parameter dependence of the critical velocity}
\label{s:parameter}

Next, we discuss the dependence of the critical velocity on the
SO-coupling strength $\kappa$ and on the obstacle radius $R$.
We focus on the moving directions $\theta_v=0$ and $\theta_v = \pi$ of the
obstacle potential.
The critical velocity $v_c$ is defined as the velocity above which the
drag force suddenly increases, as indicated by the red line in
Fig.~\ref{f:diagram}.

\begin{figure}[tb]
\includegraphics[width=9cm]{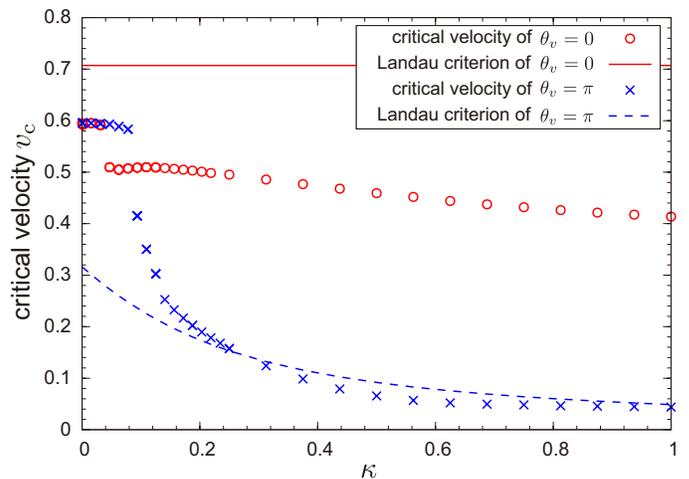}
\caption{
Dependence of the critical velocity $v_c$ on the strength of the SO
coupling $\kappa$ for $\gamma=0.8$ and $R=0.5$.
The red circles and blue crosses indicate the critical velocities for
$\theta_v=0$ and $\theta_v=\pi$, respectively.
The red solid line and the blue dashed line indicate the Landau criterion
for $\theta_v=0$ and $\theta_v=\pi$, respectively, in Eq.~(\ref{eq:SV}).
} 
\label{f:vc_kappa}
\end{figure}
Figure~\ref{f:vc_kappa} shows the $\kappa$ dependence of the critical
velocity $v_c$.
For $\theta_v=0$, $v_c$ discontinuously decreases at $\kappa \simeq 0.03$,
and gradually decreases with increasing in $\kappa$.
The discontinuous change is attributed to the $\psi_1$-$\psi_2$ symmetry
breaking in the vortices that are generated by the potential.
For $\kappa \gtrsim 0.03$, the vortex cores in the two components are
displaced from each other~\cite{M. Kato}, which can be regarded as a pair
of half-quantized vortices, whereas the vortices generated for
$\kappa \lesssim 0.03$ are the usual topological defects of the global
phase in which $\psi_1 = \psi_2$ is preserved.
Similarly, for $\theta_v=\pi$, the symmetry between the two
components is broken for $\kappa \gtrsim 0.08$, causing the sudden change
in $v_L$.
The critical velocity $v_c$ is below the Landau critical velocity $v_L$
for $\kappa \gtrsim 0.3$, which may be due to the finite size effect of
the obstacle potential.

\begin{figure}[tb]
\includegraphics[width=9cm]{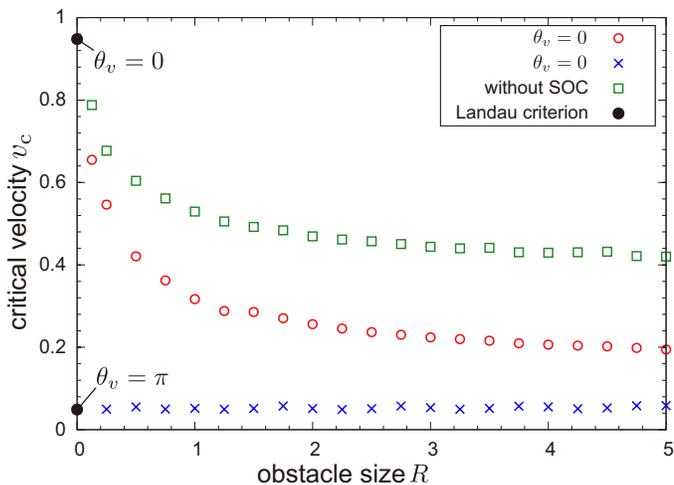}
\caption{
Dependence of critical velocities $v_c$ on the obstacle radius $R$ for
$\kappa=1$ and $\gamma=0.8$.
The red circles and blue crosses indicate the critical velocities for
$\theta_v=0$ and $\theta_v=\pi$, respectively.
The green squares indicate the critical velocity without the SOC.
The black solid circles indicate the Landau criteria in
Eq.~(\ref{eq:SV}).
}
\label{f:vc_size}
\end{figure}
Figure~\ref{f:vc_size} shows the dependence of critical velocity $v_c$ on
the obstacle radius $R$ for $\theta_v=0$ and $\theta_v=\pi$.
The critical velocity for the system without the SOC is also plotted for
comparison.
In the limit of small $R$, the critical velocities approach the Landau
critical velocity, as expected.
The critical velocity with SOC is always smaller than that without the SOC,
for the same reason as that for the steep decrease of $v_c$ in
Fig.~\ref{f:vc_kappa}, i.e., the symmetry breaking between the two
components due to the spin dependent force by the SOC tends to decrease
the critical velocity.
Interestingly, for the case of $\theta_v=\pi$, $v_c$ is approximately
independent of $R$, whereas $v_c$ decreases with $R$ for the case of
$\theta_v=0$.

\begin{figure}[tb]
\includegraphics[width=7cm]{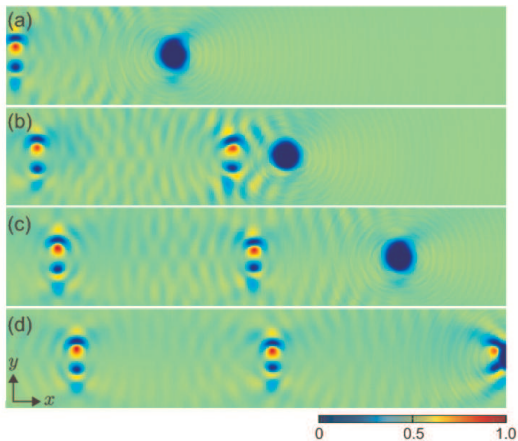}
\caption{
Serial snapshots of the density distributions $|\psi_1|^2$ of the dynamics
in the rest frame for $\kappa=1$ and $\gamma=0.8$.
The velocity of the moving obstacle is $(v,\theta_v)=(0.5,0)$, and the
radius is $R=2$.
The time interval is 75, and the field of view is $100 \times 20$.
See the Supplemental Material for movies of the dynamics of
of component 1~\cite{SM}.
}
\label{f:static}
\end{figure}
Figure~\ref{f:static} shows serial snapshots of the density distribution
in the rest frame for $(v,\theta_v)=(0.5,0)$.
The velocity of the vortex pairs released from the moving obstacle
potential, $v_{\rm vortex} \simeq 0.05$, is much smaller than that without
the SOC~\cite{M. Kato}.
In the rest frame, the vortex pairs are created just like a moving
obstacle would leave behind them on its trajectory.

\section{velocity field}
\label{s:VF}

Finally, we discuss the velocity field distribution induced by the
obstacle potential.
The velocity field of the condensate is defined as
\begin{equation}
 \label{eq:VF}
  \bm v_f(\bm r) =
  \frac{1}{2i \rho}\left[\bm{\Psi}^\dagger \bm \nabla \bm{\Psi}
  - \bm{\Psi}^T \bm \nabla \bm{\Psi}^\ast \right]
  -\frac{\kappa}{\rho}\bm{\Psi}^\dagger \bm{\sigma}_{\perp} \bm{\Psi},
\end{equation}
where $\rho = |\psi_1|^2 + |\psi_2|^2$.
The first term on the right-hand side of Eq.~(\ref{eq:VF}) is the usual
superfluid velocity, and the second term originates from the SOC.
The velocity field satisfies the equation of continuity,
$\partial \rho / \partial t + \bm \nabla \cdot (\rho \bm v) = 0$.
The velocity field far from the obstacle potential vanishes due to the
cancellation between the first and second terms of Eq.~(\ref{eq:VF}).

\begin{figure}[tb]
 \includegraphics[width=9cm]{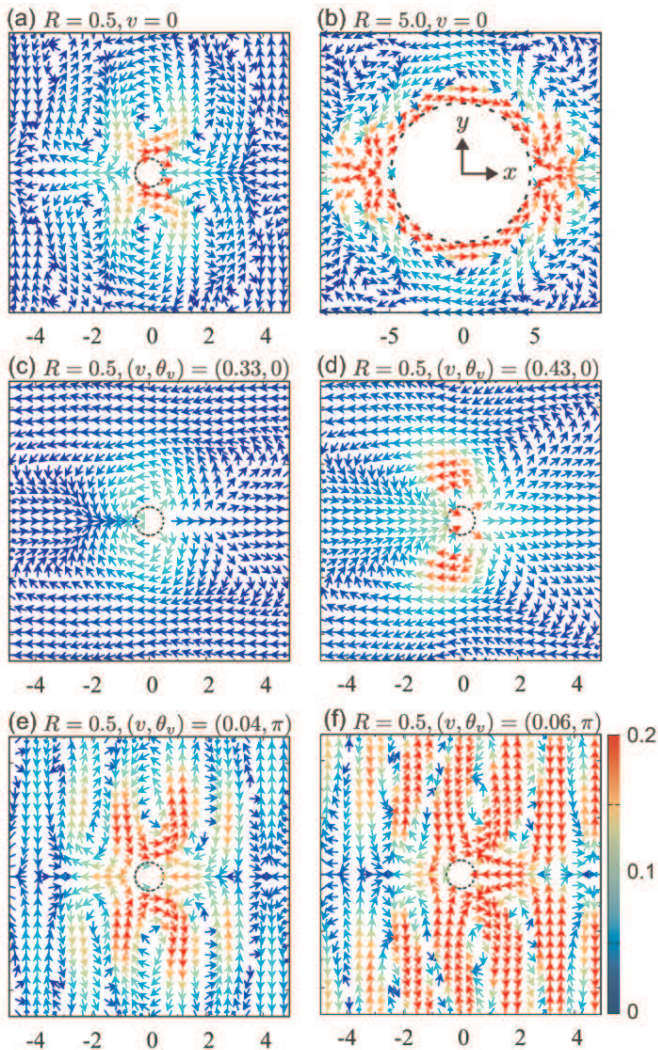}
 \caption{
 Velocity field distribution induced by an obstacle potential
 for $\kappa=1$ and $\gamma=0.8$.
 (a)-(b) Ground states with the obstacle potential at rest.
 The radii of the circular potential are $R=0.5$ and $5.0$.
 The velocity of the potential with radius $R=0.5$ is increased as
 $v(t) = 6.0\times10^{-4}t$ in (c)-(d) and $v(t) = -7.0\times10^{-5}t$ in (e)-(d).
 (c)$(v, \theta_v) = (0.33, 0)$ at $t = 550$,
 (d)$(v, \theta_v) = (0.43, 0)$ at $t = 717$,
 (e)$(v, \theta_v) = (0.04, \pi)$ at $t = 600$, and
 (f)$(v, \theta_v) = (0.06, \pi)$ at $t = 835$.
 The arrows indicate the direction of the velocity field, and their color
 indicates the value of $v$. 
 The dashed circle in each panel indicates the obstacle potential.
 }
 \label{f:VF}
\end{figure}
Figures~\ref{f:VF}(a) and \ref{f:VF}(b) show the velocity field of the
ground state with the obstacle potential at rest.
Even for the static case, the velocity field exhibits complicated
structures containing multiple circulations.
A strong rightward flow is observed in the vicinity of the potential,
which is explained as follows.
The spin-dependent SOC forces on components 1 and 2 are in the $+y$ and
$-y$ directions, respectively, which results in the density difference
between the two components at the edge of the potential.
For the wave function,
\begin{equation}
 \bm \Psi(\bm r) =
 \left(
  \begin{array}{c}
   u_1(\bm r) e^{i \kappa x}\\
   u_2(\bm r) e^{i \kappa x}
  \end{array}
   \right),
\end{equation}
with real functions $u_1$ and $u_2$, the velocity field is given by
$\bm v_f = \kappa \rho^{-1} (u_1 - u_2)^2 \bm{e}_x$, where $\bm{e}_x$
is the unit vector in the $x$ direction.
Thus, the density imbalance at the edge of the potential generates the
rightward velocity field.

For $\theta_v=0$, as the obstacle velocity increases, the circulations in
the velocity field vanish and the velocity field becomes similar to
that for the system without SOC, as shown in Figs.~\ref{f:VF}(c) and
\ref{f:VF}(d).
The disappearance of the complicated velocity field is due to the
disappearance of the imbalance between $|\psi_1|^2$ and $|\psi_2|^2$ by
the fast motion of the obstacle.

On the other hand, the velocity field for $\theta_v=\pi$ is quite
different from the case of $\theta_v=0$, as shown in Figs.~\ref{f:VF}(e)
and \ref{f:VF}(f). 
In this case, the flows are mainly along the $y$ direction, and
alternately shift upward and downward.
This behavior can be understood based on the results of the Bogoliubov
analysis in Sec. \ref{s:bogoliubov}.
Assuming $\gamma = 1$, for simplicity, the most unstable wave number is
estimated to be $\bm q \simeq -2\kappa \bm e_x$ from Eq.~(\ref{qx}), and
the eigenvector in Eq.~(\ref{eq:EV_pi}) is approximated to be $\bm{\alpha}
\propto (1, -1)^T$ and $\bm{\beta} \simeq \bm{0}$.
Substituting these into Eq.~(\ref{eq:VWF}), the excited wave function
becomes
\begin{equation}
 \Psi(\bm r,t)=
  \frac{e^{-i\mu t}}{\sqrt{2}}
  \left[
   \left(
    \begin{array}{c}
     1\\
     1
    \end{array}
   \right) e^{i\kappa x}
   + \delta u
   \left(
    \begin{array}{c}
     1\\
     -1
    \end{array}
   \right) e^{-i\kappa x}
  \right],
\end{equation}
where $\delta u$ is an infinitesimal amplitude of the excitation, which is
assumed to be real without loss of generality.
The velocity field in Eq.~(\ref{eq:VF}) is thus obtained as
\begin{equation}
 \label{eq:VF_BA}
 \bm v_f(\bm r) = -2 \delta u \sin(2\kappa x)\bm e_y,
\end{equation}
where $\bm e_y$ is a unit vector in the $y$ direction.
In Fig.~\ref{f:VF}(f), the wave length of the velocity field is estimated
to be $\lambda_v \simeq 3$, which agrees well with the wavelength $\pi /
\kappa$ in Eq.~(\ref{eq:VF_BA}).

\section{Conclusions}
\label{s:conclusions}

In conclusion, we have investigated the dynamics of an SO-coupled BEC
with a moving obstacle potential.
We found that the obstacle potential moving in the plane-wave state
exhibits a variety of excitation dynamics.
We have shown that the dynamics strongly depend on the direction of the
obstacle motion.
When the potential moves in the plane-wave direction, half-quantized
vortex pairs are released.
When the potential moves in the opposite direction, on the other hand,
spin waves are dominant.
This behavior can be understood from the Bogoliubov spectrum.
Although the Landau critical velocity derived from the Bogoliubov spectrum
is zero for the other directions, we numerically found that there is an
effective critical velocity, below which excitation is negligible and
above which the drag force on the obstacle increases steeply.
We obtained a diagram of the excitation behavior with respect to the
velocity $\bm{v}$ of the potential.
We explored the dependence of the effective critical velocity on the
strength of the SO coupling $\kappa$ and the obstacle size $R$.
We also investigated the velocity field distribution for the system,
which exhibits complicated flow patterns depending on the parameters.

\begin{acknowledgments}
The present study was supported by JSPS KAKENHI Grant Numbers JP25103007,
JP16K05505, JP17K05595, and JP17K05596, by the key project fund of the CAS for the ``Western Light'' Talent Cultivation Plan under Grant No. 2012ZD02, and by the Youth Innovation Promotion Association of CAS under Grant No. 2015334.
\end{acknowledgments}

\end{document}